\documentclass[11pt]{article}
\usepackage{graphicx}
\usepackage{latexsym,graphicx}
\usepackage{float}
\usepackage{amsmath}
\usepackage{amsfonts}
\usepackage{amssymb}
\usepackage{epstopdf}
\usepackage{color}
\DeclareGraphicsExtensions{.eps}
\usepackage[left=2.5cm,top=2.5cm,right=2.5cm,bottom=2.5cm]{geometry}

\catcode `\@=11
\catcode `\@=12
\begin{document}
	\begin{center}
		\large{\bf{A flat FLRW dark energy model in $f(Q, C)$-gravity theory with observational constraints}} \\
		\vspace{10mm}
		\normalsize{ Anirudh Pradhan$^1$, Archana Dixit$^2$, M. Zeyauddin$^3$, S. Krishnannair$^4$  }\\
		\vspace{5mm}
		
		\normalsize{$^{1}$Centre for Cosmology, Astrophysics and Space Science (CCASS), GLA University, Mathura-281 406, Uttar Pradesh, India} \\
		\vspace{2mm}
		\normalsize{$^{2}$Department of Mathematics, Institute of Applied Sciences and Humanities, GLA University, Mathura-281 406, Uttar Pradesh, India}\\ 
			\vspace{2mm}
		\normalsize{$^{3}$Department of General Studies (Mathematics)
			Jubail Industrial College Jubail 31961, Saudi Arabia}\\ 
			\vspace{2mm}
		\normalsize{$^{4}$Department of Mathematical Sciences, University of Zululand Private Bag X1001 Kwa-Dlangezwa 3886 South Africa}\\
		\vspace{2mm}
		$^1$E-mail:pradhan.anirudh@gmail.com\\
		\vspace{2mm}
		$^2$E-mail:archana.dixit@gla.ac.in \\
		\vspace{2mm}
		$^3$E-mail:uddin m@rcjy.edu.sa\\
		\vspace{2mm}
		$^{4}$E-mail:KrishnannairS@unizulu.ac.za  \\
		
		\vspace{10mm}
		
	\end{center}
	\vspace{5mm}

	\begin{abstract}
		In the recently suggested modified non-metricity gravity theory with boundary term in a flat FLRW spacetime universe, dark energy scenarios of cosmological models are examined in this study. An arbitrary function, $f(Q,C)=Q+\alpha C^{2}$, has been taken into consideration, where $Q$ is the non-metricity scalar, $C$ is the boundary term denoted by $C=\mathring{R}-Q$, and $\alpha$ is the model parameter, for the action that is quadratic in $C$. The Hubble function $H(z)=H_{0}[c_{1}(1+z)^{n}+c_{2}]^{\frac{1}{2}}$, where $H_{0}$ is the current value of the Hubble constant and $n, c_{1}$ and $c_{2}$ are arbitrary parameters with $c_{1}+c_{2}=1$, has been used to examine the dark energy characteristics of the model. We discovered a transit phase expanding universe model that is both decelerated in the past and accelerated in the present, and we discovered that the dark energy equation of state (EoS) $\omega^{(de)}$ behaves as $(-1\le\omega^{(de)}<2)$. The Om diagnostic analysis reveals the quintessence behavior in the present and the cosmological constant scenario in the late-time universe. Finally, we calculated the universe's current age, which was found to be quite similar to recent data.
	\end{abstract}
	\smallskip
	\vspace{5mm}
	\vspace{1cm}
	
	\section{Introduction}
	
	Numerous observational observations \cite{ref1}-\cite{ref4} support the so-called $\Lambda$CDM model, which is the established theory of cosmology. They depict a universe that is only now beginning to expand at an accelerated rate. The easiest way to understand dark energy within the context of general relativity (GR), which is responsible for the late-time acceleration, is through the cosmological constant $\Lambda$, with a negative-constant equation of state \cite{ref5}. Despite Lambda's remarkable effectiveness in describing current data, its fundamental interpretation is still far from being understood, largely because of the fine-tuning problem inherent in the energy of the vacuum \cite{ref6}. Modified gravity theories, such as $f(R)$, $f(T)$, and $f(Q)$ theories \cite{ref7}-\cite{ref11}, have drawn a lot of attention recently among all potential explanations for the effects of dark energy (see, for example, \cite{ref12,ref13}).\\
	
	A fascinating possibility that has just lately been researched is the gravitational interaction mediated by non-metricity, when curvature and torsion are disappearing \cite{ref14,ref15,ref16}. Since gravity can be regarded as a gauge theory without explicitly assuming the validity of the Equivalence Principle, this approach can be vital for understanding gravity at its most fundamental level. The cosmic speed increase that results from the intrinsic implications of a different geometry than the Riemannian one can be viewed from new angles by looking into the $f(Q)$ theories, where $Q$ is the non-metricity scalar. A power-law function is assumed in the analysis of the connecting matter in $f(Q)$ gravity by \cite{ref17} and a model-independent reconstruction approach is recently employed to investigate cosmological characteristics in \cite{ref18}. Nonmetricity formulation of general relativity and its scalar-tensor extension is presented in \cite{ref19} and General relativity with spin and torsion is investigated in \cite{ref20}. A Covariant formulation of $f(Q)$ theory is given in \cite{ref21} and an extension of $f(Q)$ gravity in the form of $f(Q, T)$ theory
	is proposed in \cite{ref22}.
	
	The above-mentioned novel classes of modified gravity exist in the curvature, torsional, and non-metricity situations, despite similarities between the non-modified theories at the level of equations. The reason for this is that when the torsion scalar $T$ and the non-metricity scalar $Q$ are used in place of the standard Levi-Civita Ricci scalar $\mathring{R}$ of general relativity, arbitrary functions $f(\mathring{R})$, $f(T)$, and $f(Q)$ no longer differ by a complete derivative. Recently, a new extension theory of $f(Q)$ gravity is proposed in \cite{ref23} by adding the boundary terms $C=\mathring{R}-Q$ and is called as $f(Q,C)$ gravity theory. Recently, the role of boundary terms in $f(Q)$ theory is investigated by \cite{ref24} in details. These initial studies suggest that the boundary term may behaves role of dark energy factor. Therefore, motivated by above newly proposed theory, in this paper, we will investigate behaviour of dark energy cosmological models in $f(Q,C)$ gravity with perfect-fluid.\\

 Here, we analyze the dynamical dark energy behavior of dark energy models in the current research by investigating the EoS parameters $\omega$ in various contexts by employing a particular Hubble function, which was inspired by the aforementioned studies. For this, a Lagrangian action with the formula $L=\frac{1}{16\pi}[f(Q,C)+16\pi L_{m}]\sqrt{-g}$, and any arbitrary function $f(Q,C)=Q+\alpha C^{2}$ have been taken into consideration. The $f(Q,C)$ gravity formulation was briefly described in section 2, and the field equations were derived in section 3. Section 4 displays cosmological solutions using a specific Hubble function, while Section 5 examines observational constraints using three datasets, $H(z)$, Union 2.1 and Bined datasets of SNe Ia. The result analysis and comparison with various cosmological parameters are covered in section 6, and finally conclusions are mentioned in section 7.

\section{$f(Q,C)$ Gravity}

To examine the cosmological properties of nonmetric gravity, let's start with the most general form of the affine connections \cite{ref19}
\begin{equation}\label{eq1}
  {\Gamma^{\lambda}}_{\mu\nu}={\mathring{\Gamma}^{\lambda}}_{\mu\nu}+{K^{\lambda}}_{\mu\nu}+{L^{\lambda}}_{\mu\nu}
\end{equation}
The metric tensor $g_{\mu\nu}$'s Levi-Civita connection is provided here by
\begin{equation}\label{eq2}
 {\mathring{\Gamma}^{\lambda}}_{\mu\nu}\equiv\frac{1}{2}g^{\lambda\beta}(\partial_{\mu}g_{\beta\nu}+\partial_{\nu}g_{\beta\mu}-\partial_{\beta}g_{\mu\nu}),
\end{equation}
whereas
\begin{equation}\label{eq3}
 {K^{\lambda}}_{\mu\nu}\equiv\frac{1}{2}g^{\lambda\beta}(\mathcal{T}_{\mu\beta\nu}+\mathcal{T}_{\nu\beta\mu}+\mathcal{T}\beta\mu\nu)
\end{equation}
\begin{equation}\label{eq4}
  {L^{\lambda}}_{\mu\nu}\equiv\frac{1}{2}g^{\lambda\beta}(-Q_{\mu\beta\nu}-Q_{\nu\beta\mu}+Q_{\beta\mu\nu})
\end{equation}
are, respectively, the disformation and contortion tensors. The torsion tensor, ${\mathcal{T}^{\lambda}}_{\mu\nu}\equiv{\Gamma^{\lambda}}_{\mu\nu}-{\Gamma^{\lambda}}_{\nu\mu}$, and the non-metricity tensor, given by
\begin{equation}\label{eq5}
  Q_{\rho\mu\nu}\equiv\nabla_{\rho}g_{\mu\nu}=\partial_{\rho}g_{\mu\nu}-{\Gamma^{\beta}}_{\rho\mu}g_{\beta\nu}-{\Gamma^{\beta}}_{\rho\nu}g_{\mu\beta}.
\end{equation}
Thus, some choices made on the linkages will determine the properties of the metric-affine spacetime. Torsion and curvature are assumed to be disappearing in our analysis, leading to non-metricity as the source of geometry. The non-metricity tensor is represented by two different traces, ${Q_{\mu}={Q_{\mu}}^{\alpha}}_{\alpha}$ and $\tilde{Q}^{\mu}={Q_{\alpha}}^{\mu\alpha}$, depending on the contraction order. The non-metricity scalar can therefore be defined as \cite{ref11}.

\begin{equation}\label{eq6}
  Q=-\frac{1}{4}Q_{\alpha\beta\mu}Q^{\alpha\beta\mu}+\frac{1}{2}Q_{\alpha\beta\mu}Q^{\beta\mu\alpha}+\frac{1}{4}Q_{\alpha}Q^{\alpha}-\frac{1}{2}Q_{\alpha}\tilde{Q}^{\alpha}.
\end{equation}
which, for general diffeomorphisms, is an invariant quadratic combination.\\
Last but not least, the relations can be further obtained by applying the torsion-free and curvature-free constraints (all values with a $\mathring{()}$ are calculated with respect to the Levi-Civita connection $\mathring{\Gamma}$):
\begin{equation}\label{eq7}
  \mathring{R}_{\mu\nu}+\mathring{\nabla}_{\alpha}{L^{\alpha}}_{\mu\nu}-\mathring{\nabla}_{\nu}\tilde{L}_{\mu}+\tilde{L}_{\alpha}{L^{\alpha}}_{\mu\nu}-L_{\alpha\beta\nu}{L^{\alpha\beta}}_{\mu}=0,
\end{equation}
\begin{equation}\label{eq8}
  \mathring{R}+\mathring{\nabla}_{\alpha}(L^{\alpha}-\tilde{L}^{\alpha})-Q=0.
\end{equation}
Thus, it is evident that since $Q^{\alpha}-\tilde{Q}^{\alpha}=L^{\alpha}-\tilde{L}^{\alpha}$, we consider the boundary term from the previous relation \cite{ref17, ref18}.

\begin{equation}\label{eq9}
  C=\mathring{R}-Q=-\mathring{\nabla}_{\alpha}(Q^{\alpha}-\tilde{Q}^{\alpha})=-\frac{1}{\sqrt{-g}}\partial_{\alpha}[\sqrt{-g}(Q^{\alpha}-\tilde{Q}^{\alpha})].
\end{equation}
Now, we consider the following action for $f(Q,C)$ gravity theory \cite{ref23,ref24}
\begin{equation}\label{eq10}
  S=\int{\frac{1}{16\pi}[f(Q,C)+16\pi \mathcal{L}_{m}]\sqrt{-g}d^{4}x}
\end{equation}
where $f(Q,C)$ is an arbitrary function of non-metricity scalar $Q$ and $C$ is the boundary term, and $L_{m}$ denotes the matter Lagrangian and $g$ is the determinant of the metric tensor $g_{\mu\nu}$.\\

Field equations emerge from variations in the action with respect to the metric as follows:
\begin{equation}\label{eq11}
  \frac{2}{\sqrt{-g}}\partial_{\lambda}(\sqrt{-g}f_{Q}{P^{\lambda}}_{\mu\nu})+(P_{\mu\alpha\beta}{Q_{\nu}}^{\alpha\beta}-2P_{\alpha\beta\nu}{Q^{\alpha\beta}}_{\mu})f_{Q}-\frac{f}{2}g_{\mu\nu}+\left(\frac{C}{2}g_{\mu\nu}-\mathring{\nabla}_{\mu}\mathring{\nabla}_{\nu}+g_{\mu\nu}\mathring{\nabla}^{\alpha}\mathring{\nabla}_{\alpha}-2{P^{\lambda}}_{\mu\nu}\partial_{\lambda}\right)f_{C}=8\pi T_{\mu\nu},
\end{equation}
where $f_{Q}=\frac{\partial{f}}{\partial{Q}}$ and $f_{C}=\frac{\partial{f}}{\partial{C}}$, and where $T_{\mu\nu}$ is the matter energy momentum tensor. We observe here that by altering the action with respect to the affine connection, we may obtain the connection field equation because the affine connection is independent of the metric tensor \cite{ref23}.

\begin{equation}\label{eq12}
  (\nabla_{\mu}-\tilde{L}_{\mu})(\nabla_{\nu}-\tilde{L}_{\nu})[4(f_{Q}-f_{C}){P^{\mu\nu}}_{\lambda}+{\Delta_{\lambda}}^{\mu\nu}]=0
\end{equation}
where ${\Delta_{\lambda}}^{\mu\nu}=-\frac{2}{\sqrt{-g}}\frac{\delta(\sqrt{-g}\mathcal{L}_{m})}{\delta{\Gamma^{\lambda}}_{\mu\nu}}$ is the hyper momentum tensor \cite{ref22}. As
\begin{equation}\label{eq13}
  \partial_{\nu}\sqrt{-g}=-\sqrt{-g}\tilde{L}_{\nu}.
\end{equation}
The previous connection field equation can be re-expressed as while taking the coincident gauge as
\begin{equation}\label{eq14}
  \partial_{\mu}\partial_{\nu}(\sqrt{-g}[4(f_{Q}-f_{C}){P^{\mu\nu}}_{\lambda}+{\Delta_{\lambda}}^{\mu\nu}])=0,
\end{equation}
if this is the case, the field equation that is equivalent to the $f(Q)$ gravity equation will be found \cite{ref17}.\\
The second and third terms on the right-hand side make up the $f(Q)$ theory, as we can see from the field equation \eqref{eq11}. Following the conventional computation (for an example, see \cite{ref21,ref22}), we arrive at

\begin{equation}\label{eq15}
  \frac{2}{\sqrt{-g}}\partial_{\lambda}(\sqrt{-g}f_{Q}{P^{\lambda}}_{\mu\nu})+(P_{\mu\alpha\beta}{Q_{\nu}}^{\alpha\beta}-2P_{\alpha\beta\nu}{Q^{\alpha\beta}}_{\mu})f_{Q}=\left(\frac{Q}{2}g_{\mu\nu}+\mathring{G}_{\mu\nu}+2{P^{\lambda}}_{\mu\nu}\partial_{\lambda}\right)f_{Q},
\end{equation}
where the Einstein tensor associated with the Levi-Civita link is $\mathring{G}_{\mu\nu}$. Consequently, we may covariantly rewrite the metric field equation as

\begin{equation}\label{eq16}
-\frac{f}{2}g_{\mu\nu}+2{P^{\lambda}}_{\mu\nu}\nabla_{\lambda}(f_{Q}-f_{C})+\left(\frac{Q}{2}g_{\mu\nu}+\mathring{G}_{\mu\nu}\right)f_{Q}+\left(\frac{C}{2}g_{\mu\nu}-\mathring{\nabla}_{\mu}\mathring{\nabla}_{\nu}+g_{\mu\nu}\mathring{\nabla}^{\alpha}\mathring{\nabla}_{\alpha}\right)f_{C}=8\pi T_{\mu\nu}.
\end{equation}
The effective stress energy tensor is defined as follows:
\begin{equation}\label{eq17}
  T^{eff}_{\mu\nu}=T_{\mu\nu}+\frac{1}{8\pi}\left[\frac{f}{2}g_{\mu\nu}-2{P^{\lambda}}_{\mu\nu}\nabla_{\lambda}(f_{Q}-f_{C})-\frac{Qf_{Q}}{2}g_{\mu\nu}-\left(\frac{C}{2}g_{\mu\nu}-\mathring{\nabla}_{\mu}\mathring{\nabla}_{\nu}+g_{\mu\nu}\mathring{\nabla}^{\alpha}\mathring{\nabla}_{\alpha}\right)f_{C}\right]
\end{equation}
and as a result, we get
\begin{equation}\label{eq18}
  \mathring{G}_{\mu\nu}=\frac{8\pi}{f_{Q}} T^{eff}_{\mu\nu}
\end{equation}
As a result, we acquire an additional, effective sector of geometrical origin within the context of $f(Q,C)$ gravity.\\

If the function $f$ is linear in $C$, then \eqref{eq16} reduces to the standard field equation for $f(Q)$ gravity: $f(Q)=f(Q) + \beta C$.

\begin{equation}\label{eq19}
-\frac{f}{2}g_{\mu\nu}+2{P^{\lambda}}_{\mu\nu}\nabla_{\lambda}f_{Q}+\left(\frac{Q}{2}g_{\mu\nu}+\mathring{G}_{\mu\nu}\right)f_{Q}=8\pi T_{\mu\nu}.
\end{equation}

\section{Field Equations}

In this part, $f(Q,C)$ cosmology is introduced, and $f(Q,C)$ gravity is applied to a cosmological framework. We study the homogeneous and isotropic flat Friedmann-Robertson-Walker (FRW) spacetime given by the line element in Cartesian coordinates.

\begin{equation}\label{eq20}
  ds^{2}=-dt^{2}+a^{2}(t)[dx^{2}+dy^{2}+dz^{2}],
\end{equation}
where the scale factor $a(t)$ is given by its first time derivative, which yields the Hubble parameter $H(t)=\frac{\dot{a}}{a}$. The corresponding stress-energy momentum tensor is given by
\begin{equation}\label{eq21}
	T_{\mu\nu}=(\rho^{(m)}+p^{(m)})u_{\mu}u_{\nu}+p^{(m)}g_{\mu\nu}.
\end{equation}
where $\rho^{(m)}, p^{(m)}$ are matter energy density and matter fluid pressure, respectively,  and $u_{\mu} u^{\mu}=-1$ and $u^{\mu}=(0,0,0,-1)$ is the four-velocity vectors, $g_{\mu\nu}$ is the metric tensor.\\
As we said in the previous section, the framework of $f(Q,C)$ gravity enables us to obtain an additional, useful sector of geometrical origin, as shown in \eqref{eq17}. As a result, in a cosmic context, this phrase will be comparable to an effective dark-energy sector with an energy-momentum tensor.

\begin{equation}\label{eq22}
  T^{(de)}_{\mu\nu}=\frac{1}{f_{Q}}\left[\frac{f}{2}g_{\mu\nu}-2{P^{\lambda}}_{\mu\nu}\nabla_{\lambda}(f_{Q}-f_{C})-\frac{Qf_{Q}}{2}g_{\mu\nu}-\left(\frac{C}{2}g_{\mu\nu}-\mathring{\nabla}_{\mu}\mathring{\nabla}_{\nu}+g_{\mu\nu}\mathring{\nabla}^{\alpha}\mathring{\nabla}_{\alpha}\right)f_{C}\right]
\end{equation}
In this paper, we consider the case with vanishing affine connection ${\Gamma^{\alpha}}_{\mu\nu}=0$. Hence, one can obtain
\begin{equation}\label{eq23}
	\mathring{G}_{\mu\nu}=-(3H^{2}+2\dot{H})h_{\mu\nu}+3H^{2}u_{\mu}u_{\nu},
\end{equation}
\begin{equation}\label{eq24}
	\mathring{R}=6(2H^{2}+\dot{H}),
\end{equation}
\begin{equation}\label{eq25}
	Q=-6H^{2},
\end{equation}\label{eq26}
\begin{equation}
	C=\mathring{R}-Q=6(3H^{2}+\dot{H}),
\end{equation}

where $h_{\mu\nu}=g_{\mu\nu}+u_{\mu}u_{\nu}$, and $u_{\nu}=(dt)_{\nu}$. We get the Friedmann-like equations by incorporating these into the general field equations \eqref{eq11} as

\begin{equation}\label{eq27}
  3H^{2}=8\pi(\rho^{(m)}+\rho^{(de)})
\end{equation}
\begin{equation}\label{eq28}
  -(2\dot{H}+3H^{2})=8\pi(p^{(m)}+p^{(de)})
\end{equation}
where the energy density and pressure of the area of matter considered to be a perfect fluid are $\rho^{(m)}$ and $p^{(m)}$, respectively, and where the effective dark-energy density and pressure are given as

\begin{equation}\label{eq29}
  \rho^{(de)}=\frac{1}{8\pi}\left[3H^{2}(1-2f_{Q})-\frac{f}{2}+(9H^{2}+3\dot{H})f_{C}-3H\dot{f}_{C}\right],
\end{equation}
and
\begin{equation}\label{eq30}
  p(de)=\frac{1}{8\pi}\left[-2\dot{H}(1-f_{Q})-3H^{2}(1-2f_{Q})+\frac{f}{2}+2H\dot{f}_{Q}-(9H^{2}+3\dot{H})f_{C}+\ddot{f}_{C}\right],
\end{equation}
respectively.\\
Now, we define the energy conservation equation as
\begin{equation}\label{eq31}
	\dot{\rho}^{(m)}+3H[\rho^{(m)}+p^{(m)}]=0,
\end{equation}
and
\begin{equation}\label{eq32}
	\dot{\rho}^{(de)}+3H[\rho^{(de)}+p^{(de)}]=0.
\end{equation}
\section{Cosmological Solutions}
We have two linearly independent field equations \eqref{eq27} and \eqref{eq28} in four unknowns $H, \rho^{(m)}, p^{(m)}, f$, hence we need at least two constraints on these unknowns in order to discover accurate solutions of the field equations. First, we take the arbitrary quadratic in $C$ gravity function, $f(Q,C)$, as

\begin{equation}\label{eq33}
	f(Q,C)=Q+\alpha C^{2}
\end{equation}
where $\alpha$ is an arbitrary model free parameter.\\
Due to the non-linear complexity of its field equations and the requirement to examine the dynamical behavior of dark energy models under $f(Q, C)$ gravity, we must adopt a parametrization of either the scale factor $a(t)$ or the Hubble function $H(t)$. Model independent strategy to examine dark energy features of expanding universe models is the name of this technique. In this instance, we apply the parametrization of $H(t)$ as shown in \cite{ref25}, $H\propto H(a)$, and as a result, we take it to be:

\begin{equation}\label{eq34}
	H(a)=H_{0}\left( \frac{c_{1}}{a^{n}}+c_{2}\right)^{\frac{1}{2}} 
\end{equation}

where $H_{0}$ is the current value of the Hubble parameter $H$ and $c_{1}, c_{2}, n$ with $c_{1}+c_{2}=1$ are three positive arbitrary constants. The recent observational datasets \cite{ref1}-\cite{ref4} support this choice of Hubble function $H$ parametrization since it produces a scale factor that is consistent with an expanding universe in the transit phase (moving from a decelerating to an accelerating state). Now that we have the Hubble function defined as $H=\frac{\dot{a}}{a}$, we can rewrite equation \eqref{eq34} as

\begin{equation}\label{eq35}
	\frac{\dot{a}}{a}=H_{0}\left( \frac{c_{1}}{a^{n}}+c_{2}\right)^{\frac{1}{2}}
\end{equation}

Integrating Eq.~\eqref{eq35}, we obtain
\begin{equation}\label{eq36}
	a(t)=\left[ \frac{c_{3}e^{H_{0}c_{2}nt}-c_{1}}{c_{2}}\right]^{\frac{1}{n}} 
\end{equation}
where $c_{3}$ is an integrating constant and taking $t=0$, we obtain
\begin{equation}\label{eq37}
	a(0)=\left[\frac{c_{3}-c_{1}}{c_{2}}\right]^{\frac{1}{n}} 
\end{equation}
In standard convention, we take $a(0)=1$ and hence, we find $c_{3}=c_{1}+c_{2}=1$, that implies $c_{1}<1$.\\
Now, we express the Hubble function $H$ in terms of redshift $z$ using the relation \cite{ref26}
\begin{equation}\label{eq38}
	\frac{a_{0}}{a}=1+z,
\end{equation}
with standard convention $a_{0}=1$, as given below
\begin{equation}\label{eq39}
	H(z)=H_{0}[c_{1}(1+z)^{n}+c_{2}]^{\frac{1}{2}}
\end{equation}

Also, from Eq.~\eqref{eq38}, we can obtain the differential coefficients of Hubble function $\dot{H}$, $\ddot{H}$ and $\dddot{H}$ in terms of $z$ and diffential coefficients of $H(z)$ with respect to $z$, as 

\begin{equation}\label{eq40}
	\dot{H}=-(1+z)HH',
\end{equation}
\begin{equation}\label{eq41}
	\ddot{H}=(1+z)H^{2}H'+(1+z)^{2}HH'^{2}+(1+z)^{2}H^{2}H'',
\end{equation}
\begin{equation}\label{eq42}
	\dddot{H}=-(1+z)H^{3}H'-4(1+z)^{2}H^{2}H'^{2}-3(1+z)^{2}H^{3}H''-(1+z)^{3}HH'^{3}-4(1+z)^{3}H^{2}H'H''-(1+z)^{3}H^{3}H'''.
\end{equation}

Now, from Eq.~\eqref{eq33}, we have $f_{Q}=1$, $f_{C}=2\alpha C$ and using these expressions in Eq.~\eqref{eq29} \& \eqref{eq30}, we obtain the following expressions dark energy density $\rho^{(de)}$ and dark fluid pressure $p^{(de)}$ as:

\begin{equation}\label{eq43}
\rho^{(de)}=\frac{9\alpha H_{0}^{4}}{16\pi}\left[ (36-8n-3n^{2})c_{1}^{2}(1+z)^{2n}+(72-8n-4n^{2})c_{1}c_{2}(1+z)^{n}+36c_{2}^{2}\right] 
\end{equation}
and
\begin{equation}\label{eq44}
	p^{(de)}=-\frac{3\alpha H_{0}^{4}}{16\pi}\left[ (108-36n-33n^{2}+6n^{3})c_{1}^{2}(1+z)^{2n}+(216-36n-24n^{2}+4n^{3})c_{1}c_{2}(1+z)^{n}+108c_{2}^{2}\right] 
\end{equation}
respectively. Hence, the EoS parameter for dark energy is obtained as
\begin{equation}\label{eq45}
	\omega^{(de)}=-1-\frac{6n(n^{2}-4n-2)c_{1}^{2}(1+z)^{2n}+4n(n^{2}-3n-3)c_{1}c_{2}(1+z)^{n}}{3[(36-8n-3n^{2})c_{1}^{2}(1+z)^{2n}+(72-8n-4n^{2})c_{1}c_{2}(1+z)^{n}+36c_{2}^{2}]}
\end{equation}
Now, we define the equation of state (EoS) parameter $\omega$ for perfect fluid and dark energy source as $p^{(m)}=\omega^{(m)}\rho^{(m)}$ and $p^{(de)}=\omega^{(de)}\rho^{(de)}$, respectively, and solving Eq.~\eqref{eq31} with $\omega^{(m)}=constant$, we obtain the matter energy density as
\begin{equation}\label{eq46}
	\rho^{(m)}=\rho_{0}^{(m)}(1+z)^{3(1+\omega^{(m)})}
\end{equation}
Now, we define the effective EoS parameter $\omega^{eff}$ as
\begin{equation}\label{eq47}
	\omega^{eff}=\frac{p^{(m)}+p^{(de)}}{\rho^{(m)}+\rho^{(de)}}
\end{equation}

The deceleration parameter $q$ is derived as
\begin{equation}\label{eq48}
	q=-1+\frac{nc_{1} (1+z)^{n}}{2[c_{1}(1+z)^{n}+c_{2}]},
\end{equation}

\section{Observational Constraints}
\subsection{Hubble Function}

For the best fit values of model parameters in $H(z)$, we considered $46$ Hubble constant datasets of $H(z)$ across redshift $z$ with errors in $H(z)$ that are observed in \cite{ref27}-\cite{ref42} using the method of differential age (DA) time to time (see Table 1). For this investigation, we employed the $\chi^{2}$-test formula shown below:

\begin{equation}\nonumber
	\chi^{2}(n, c_{1}, H_{0})=\sum_{i=1}^{i=N}\frac{[(H_{ob})_{i}-(H_{th})_{i}]^{2}}{\sigma_{i}^{2}}
\end{equation}
Where $N$ denotes the total amount of data, $H_{ob},~H_{th}$, respectively, the observed and hypothesized datasets of $H(z)$ and standard deviations are displayed by $\sigma_{i}$.

\begin{figure}[H]
	\centering
	\includegraphics[width=10cm,height=8cm,angle=0]{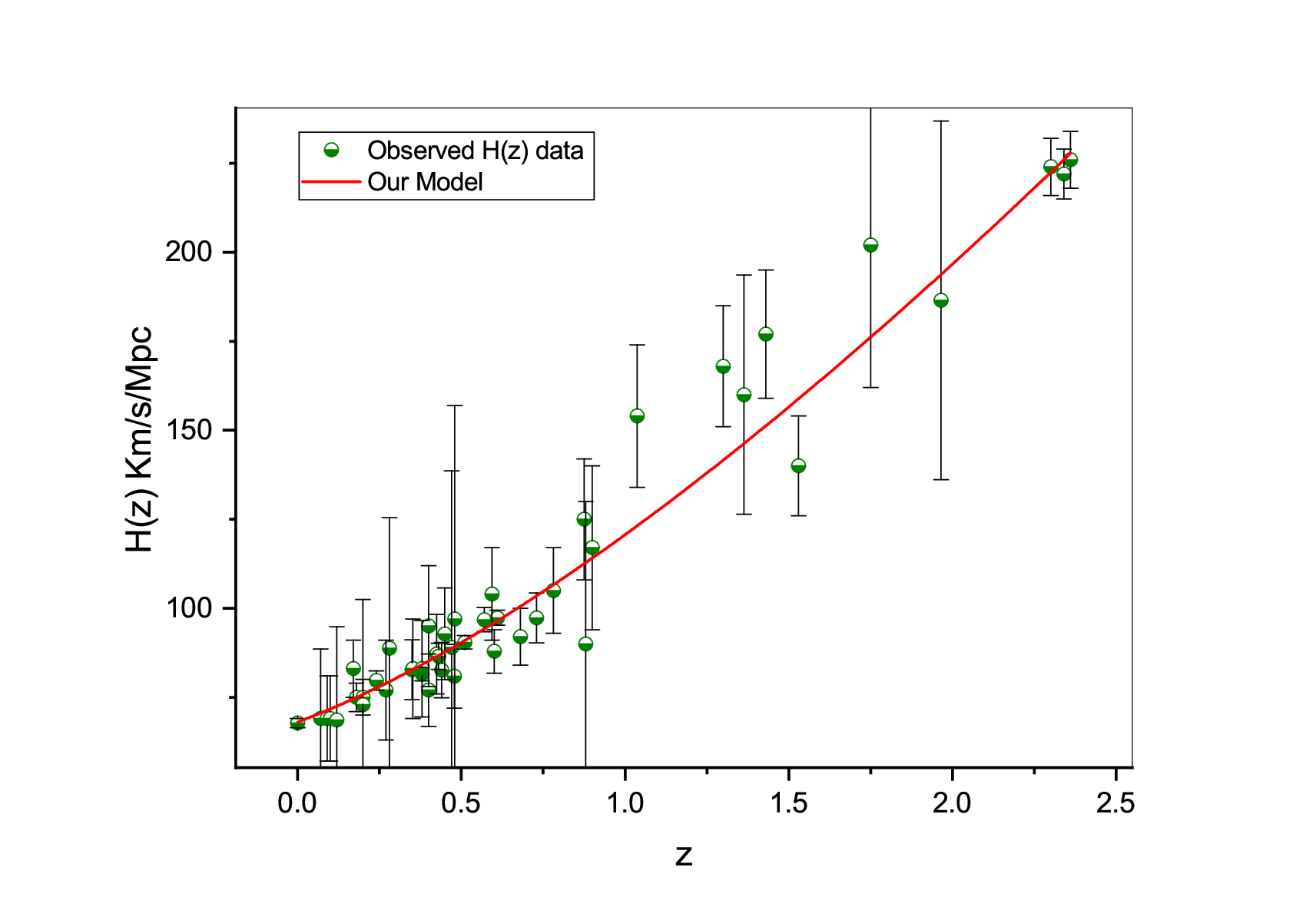}
	\caption{The best fit curve of Hubble function $H(z)$.}
\end{figure}

\begin{table}[H]
	\centering
	{\tiny \begin{tabular}{|c|c|c|c|c|c|c|c|c|c|c|c|}
			\hline
			``S.No. & $z$   & $H(z)$  & $\sigma_{H}$  & Reference   & S.No. & $z$   & $H(z)$  & $\sigma_{H}$  & Reference\\
			\hline
			1  & $0$      & $67.77$ & $1.30$   & \cite{ref27}         & 24  & $0.4783$  & $80.9$  & $9$     & \cite{ref41}\\
			2  & $0.07$   & $69$    & $19.6$   & \cite{ref28}         & 25  & $0.48$    & $97$    & $60$    & \cite{ref29}\\
			3  & $0.09$   & $69$    & $12$     & \cite{ref40}         & 26  & $0.51$    & $90.4$  & $1.9$   & \cite{ref31}\\
			4  & $0.10$   & $69$    & $12$     & \cite{ref29}         & 27  & $0.57$    & $96.8$  & $3.4$   & \cite{ref42}\\
			5  & $0.12$   & $68.6$  & $26.2$   & \cite{ref28}         & 28  & $0.593$   & $104$   & $13$    & \cite{ref39}\\
			6  & $0.17$   & $83$    & $8$      & \cite{ref29}         & 29  & $0.60$    & $87.9$  & $6.1$   & \cite{ref33}\\
			7  & $0.179$  & $75$    & $4$      & \cite{ref39}         & 30  & $0.61$    & $97.3$  & $2.1$   & \cite{ref31}\\
			8  & $0.1993$ & $75$    & $5$      & \cite{ref39}         & 31  & $0.68$    & $92$    & $8$     & \cite{ref39}\\
			9  & $0.2$    & $72.9$  & $29.6$   & \cite{ref28}         & 32  & $0.73$    & $97.3$  & $7$     & \cite{ref33}\\
			10  & $0.24$   & $79.7$  & $2.7$    & \cite{ref30}         & 33  & $0.781$   & $105$   & $12$    & \cite{ref39}\\
			11  & $0.27$   & $77$    & $14$     & \cite{ref29}         & 34  & $0.875$   & $125$   & $17$    & \cite{ref39}\\
			12  & $0.28$   & $88.8$  & $36.6$   & \cite{ref28}         & 35  & $0.88$    & $90$    & $40$    & \cite{ref29}\\
			13  & $0.35$   & $82.7$  & $8.4$    & \cite{ref32}         & 36  & $0.9$     & $117$   & $23$    & \cite{ref29}\\
			14  & $0.352$  & $83$    & $14$     & \cite{ref39}         & 37  & $1.037$   & $154$   & $20$    & \cite{ref30}\\
			15  & $0.38$   & $81.5$  & $1.9$    & \cite{ref31}         & 38  & $1.3$     & $168$   & $17$    & \cite{ref29}\\
			16  & $0.3802$ & $83$    & $13.5$   & \cite{ref32}         & 39  & $1.363$   & $160$   & $33.6$  & \cite{ref35}\\
			17  & $0.4$    & $95$    & $17$     & \cite{ref40}         & 40  & $1.43$    & $177$   & $18$    & \cite{ref29}\\
			18  & $0.004$  & $77$    & $10.2$   & \cite{ref41}         & 41  & $1.53$    & $140$   & $14$    & \cite{ref29}\\
			19  & $0.4247$ & $87.1$  & $11.2$   & \cite{ref41}         & 42  & $1.75$    & $202$   & $40$    & \cite{ref35}\\
			20  & $0.43$   & $86.5$  & $3.7$    & \cite{ref30}         & 43  & $1.965$   & $186.5$ & $50.4$  & \cite{ref30}\\
			21  & $0.44$   & $82.6$  & $7.8$    & \cite{ref33}         & 44  & $2.3$     & $224$   & $8$     & \cite{ref38}\\
			22  & $0.44497$& $92.8$  & $12.9$   & \cite{ref41}         & 45  & $2.34$    & $222$   & $7$     & \cite{ref36}\\
			23  & $0.47$   & $89$    & $49.6$   & \cite{ref34}         & 46  & $2.36$    & $226$   & $8$     & \cite{ref37}"\\
			\hline
	\end{tabular}}
	\caption{Observed values of $H(z)$.}\label{T1}
\end{table}

\begin{table}[H]
	\centering
	\begin{tabular}{|c|c|}
		\hline
		
		Parameter       & Value   \\
		\hline
		$H_{0}$         & $68\pm0.84653$\\
		$c_{1}$       & $0.36934\pm0.04197$\\
		$n$             & $2.76902\pm0.08594$\\
		Reduced $\chi^{2}$      & $0.50647$\\
		\hline
	\end{tabular}
	\caption{The best fit values of cosmological parameters in the $\chi^{2}$-test of $H(z)$ with observed $H(z)$ data with a $95\%$ confidence level of bounds.}\label{T2}
\end{table}
Table 2 lists the approximations for the cosmological parameters used in the $\chi^{2}$-test of $H(z)$, and Figure 1 depicts the best fit shape. In Eq.~\eqref{eq39}, the mathematical expression for $H(z)$ is shown. We calculated the Hubble constant ($H_{0}$) for this analysis as $H_{0}=68\pm0.84653~Km/s/Mpc$, which is in good agreement with recently measured values. 
Recently in 2018, the Plank Collaboration estimated the present value of Hubble constant $H_{0}=67.4\pm0.5$ km/s/Mpc \cite{ref43} whereas in 2021, Riess et al.\cite{ref44} have obtained as $H_{0}=73.2\pm1.3$ km/s/Mpc. For a lower $\chi^{2}$ value of $0.50647$, we calculated the best fit values of the model parameters $c_{1}=0.36934\pm0.04197$ and $n=2.76902\pm0.08594$, which are listed in Table 2.
\subsection{Apparent Magnitude $m(z)$}
One of the main observational tools for researching the evolution of the cosmos is the relationship between redshift and luminosity distance. Since the universe is expanding and distant luminous objects' light is experiencing a redshift, the luminosity distance ($D_{L}$) is calculated in terms of the cosmic redshift ($z$). It is provided as

\begin{equation}\label{eq49}
	D_{L}=a_{0} r (1+z).
\end{equation}
where the radial coordinate of the source $r$, is established by
\begin{equation}\label{eq50}
	r  =  \int^r_{0}dr = \int^t_{0}\frac{cdt}{a(t)} = \frac{1}{a_{0}}\int^z_0\frac{cdz}{H(z)}
\end{equation}
where we have used $ dt=dz/\dot{z}, \dot{z}=-H(1+z).$

As a result, the luminosity distance is calculated as follows:

\begin{equation}\label{52}
	D_{L}=c(1+z)\int^z_0\frac{dz}{H(z)}
\end{equation}
Hence, the apparent magnitude $m(z)$ of a supernova \cite{ref26,ref45,ref46} is defined as:
\begin{equation}\label{eq52}
	m(z)=16.08+ 5log_{10}\left[\frac{(1+z)H_{0}}{.026} \int^z_0\frac{dz}{H(z)}\right].
\end{equation}
We use the most recent collection of supernovae pantheon samples $40$ Bined data in the ($0.01 \le z \le 1.7$) range \cite{ref47,ref48} plus union 2.1 compilation of supernovae data in the range ($0.1 \le z \le 1.414$) \cite{ref49}. We have used the following $\chi^{2}$ formula to constrain different model parameters:
\begin{equation}\nonumber
	\chi^{2}(n, c_{1})=\sum_{i=1}^{i=N}\frac{[(m_{ob})_{i}-(m_{th})_{i}]^{2}}{\sigma_{i}^{2}}
\end{equation}
The entire amount of data is denoted by $N$, the observed and theoretical datasets of $m(z)$ are represented by $m_{ob}$ and $m_{th}$, respectively, and standard deviations are denoted by $\sigma_{i}$.\\
\begin{figure}[H]
	\centering
	a.\includegraphics[width=8cm,height=7cm,angle=0]{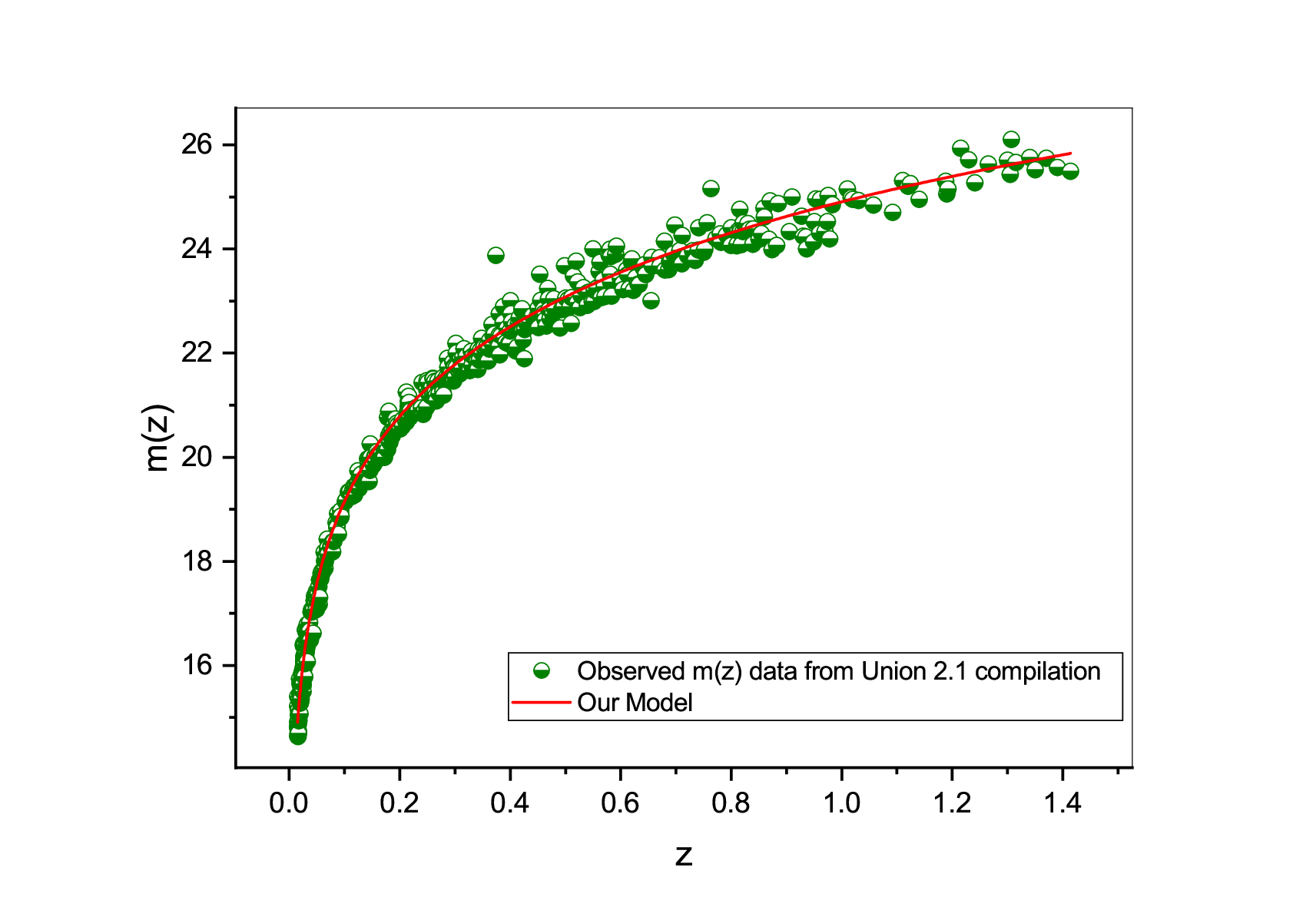}
	b.\includegraphics[width=8cm,height=7cm,angle=0]{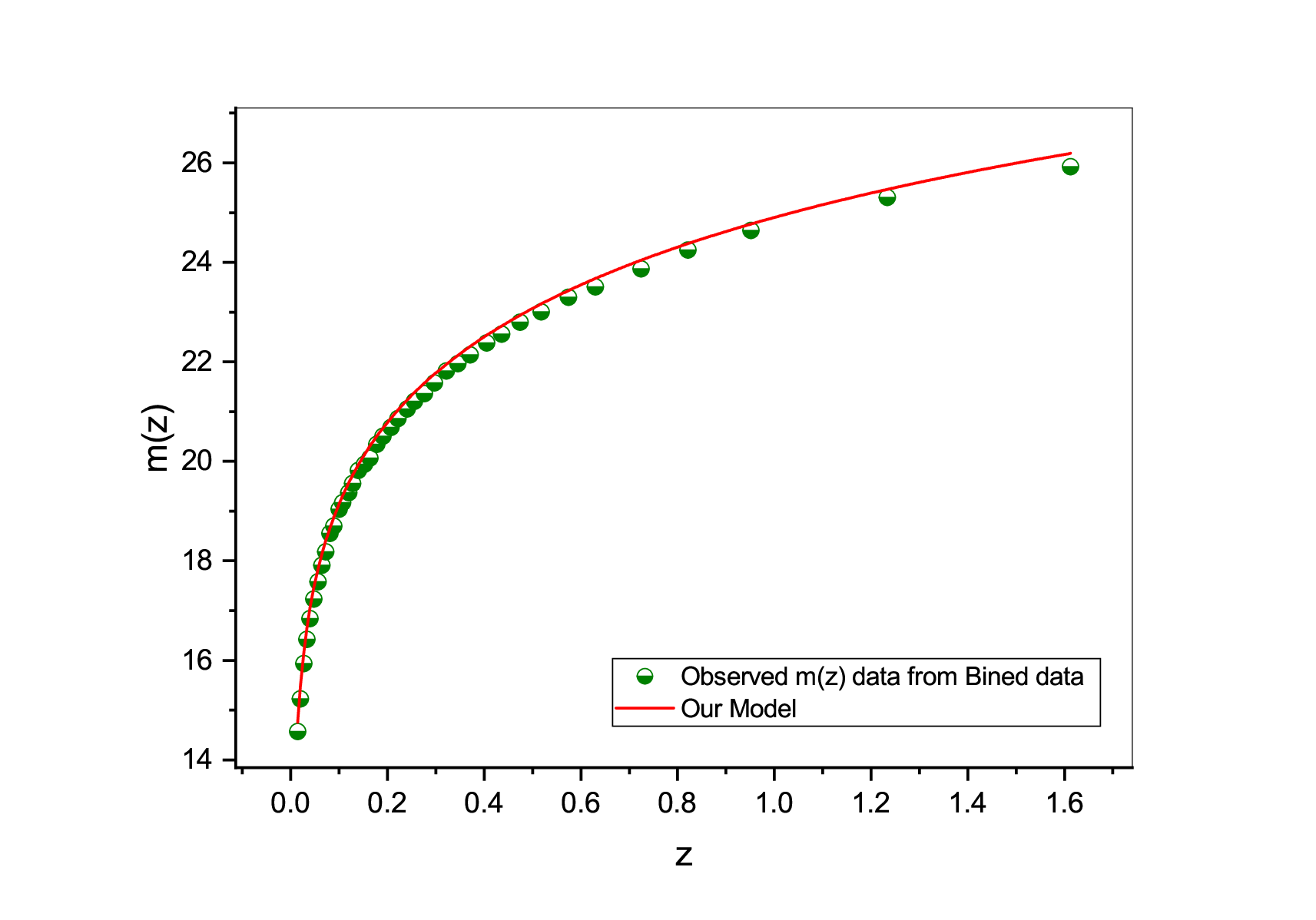}
	\caption{Best fit curve of apparent magnitude $m(z)$ over redshift $z$ with observational SNe Ia data sets, Union 2.1 and Bined, respectively.}
\end{figure}

\begin{table}[H]
	\centering
	\begin{tabular}{|c|c|c|}
		\hline
		
		Parameter             & Value for Union 2.1 data        & Value for Bined data\\
		\hline
		$c_{1}$             & $0.3670\pm0.03877$              & $0.40\pm0.032996$\\
		$n$                   & $2.83802\pm0.65173$             & $2.72294\pm1.34523$\\
		Reduced $\chi^{2}$    & $0.03497$                       & $0.05498$\\
		\hline
	\end{tabular}
	\caption{The best fit values of cosmological parameters in the $\chi^{2}$-test of $m(z)$ with observational data from Union 2.1 and Bined data at $95\%$ confidence level of bounds.}\label{T3}
\end{table}

The best fit shapes of apparent magnitude $m(z)$ over $z$ are shown in Figure 2a, 2b and its expression is given in Eq.~\eqref{eq52}. The best fit estimated values of various cosmological parameters in $\chi^{2}$-test of $m(z)$ with SNe Ia data are given in Table 3. 
\section{Result Analysis and Discussions}

\subsection*{Deceleration parameter}
Figure 3 depicts the geometric evolution of the deceleration parameter $q(z)$, and its mathematical expression is represented in equation \eqref{eq48}. Figure 3 illustrates how $q(z)$ evolves across $(-1\le z \le 3)$ and exhibits a signature-flipping point (transition point). $q(z)$ is an increasing function of $z$. Our universe model is now in an accelerating phase, according to the estimated values of the deceleration parameter along the three observational datasets $H(z)$, Union $2.1$, and Bined, respectively, which are $q_{0}=-0.4886, -0.4792, -0.4534$. We can observe that as $z\to-1$, $q\to-1$ that reveals the accelerating scenarios of late-time universe and as $z\to3$, the $q$ tends to a positive value that reveals the early decelerating scenarios of the expanding universe. Also, from Eq.~\eqref{eq48}, we can find as $z\to\infty$, $q\to-1+\frac{n}{2}$, which reveals that early universe is decelerating for $n>2$, and accelerating for $n<2$. And also, one can see that as $z\to0$, $q\to-1+\frac{nc_{1}}{2}$, that shows the accelerating current universe for $n<\frac{2}{c_{1}}$ and decelerating current universe for $n>\frac{2}{c_{1}}$. But the recent observations \cite{ref1}-\cite{ref4} suggests the existence of an universe which is decelerating in past and accelerating in current and future, and therefore, we can say that $2<n<\frac{2}{c_{1}}$ and this implies that $c_{1}$ should be $c_{1}<1$ which supports to our values obtained in observational constraints in Table 2, 3. The transition point is generally denoted by $z_{t}$ and expressed as
\begin{equation}\label{eq53}
	z_{t}=\left[ \frac{c_{2}}{(n-2)c_{1}}\right]^{\frac{1}{n}}-1
\end{equation}

\begin{figure}[H]
	\centering
	\includegraphics[width=10cm,height=8cm,angle=0]{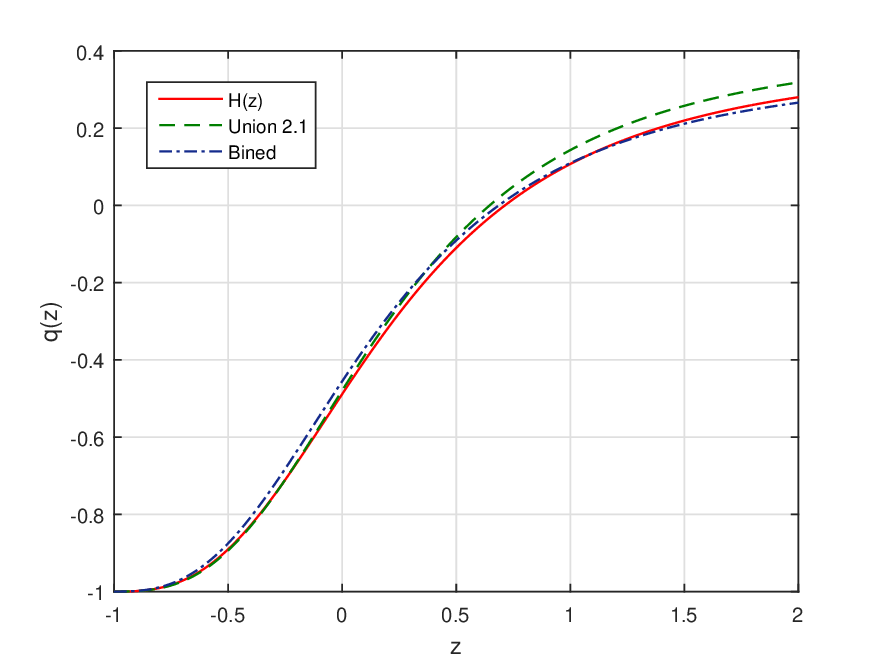}
	\caption{Variation of deceleration parameter $q(z)$ versus $z$.}
\end{figure}
According to three observational datasets, the estimated transition values are $z_{t}=0.7132, 0.6463, 0.6864$, respectively, which are quite close to the most recent observed values. The expansion phase transition point is reached at $z=z_{t}$, $q=0$, when the universe is expanding in an accelerating phase for $-1\le z<z_{t}$ and a decelerating phase for $z=z_{t}$, $q=0$.
 The ``High-z Supernova Search (HZSNS) team" recently discovered $z_{t}=0.46\pm0.13$ \cite{ref2,ref3,ref50} in 2004 and $z_{t}=0.43\pm0.07$ \cite{ref50} in 2007. The transition redshift determined by \cite{ref26,ref51} from the ``Supernova Legacy Survey (SNLS)" is $z_{t}\sim 0.6$, which is more consistent with the flat $\Lambda$CDM model ($z_{t} = (2\Omega_{\Lambda}/\Omega_{m})^{\frac{1}{3}} - 1 \sim 0.66$). $0.60 \leq z_{t}\leq 1.18$, in \cite{ref52} is an extra limitation.
\subsection*{Dark energy density and pressure}

Dark energy $\rho^{(de)}$ and dark energy pressure $p^{(de)}$ are mathematically expressed as Eqs.~\eqref{eq43} and \eqref{eq44}, respectively, and their geometrical behavior is depicted in Figures 4a and 4b, respectively. Figure 4a shows that $\rho^{(de)}$ is an increasing function of $z$ over $(-1, 0.4)$ and that its value drops as redshift increases from $z\ge0.4$. According to Figure 4b, $p^{(de)}$ is an increasing function of both $z$ and $p^{(de)}<0$ at the present and late stages of its evolution. This negative pressure in the dark energy component of the cosmos is what drives the acceleration of the expanding universe. From Eqs.~\eqref{eq43} \& \eqref{eq44}, we may deduce that the density and pressure of dark energy are given by
\begin{equation*}
	\rho^{(de)}\to\frac{81\alpha c_{2}^{2}H_{0}^{4}}{4\pi},~~~~p^{(de)}\to-\frac{81\alpha c_{2}^{2}H_{0}^{4}}{4\pi}.
\end{equation*}
Here, we have chosen the value of model parameter $\alpha$ for which $\rho^{(de)}=\Lambda$ and this this gives $\alpha=\frac{4\pi\Lambda}{81c_{2}^{2}H_{0}^{4}}$, where we have used the value of $\Lambda=2.036\times10^{-35} s^{-2}$ which is supported by \cite{ref5}-\cite{ref7}, the characteristics of dark energy and pressure, cause of acceleration in expanding universe. Thus, at late-time universe $\rho^{(de)}\to\Lambda$ and $p^{(de)}\to-\Lambda$ which reveals the consistency of our universe model with $\Lambda$CDM model.
\begin{figure}[H]
	\centering
	a.\includegraphics[width=8cm,height=7cm,angle=0]{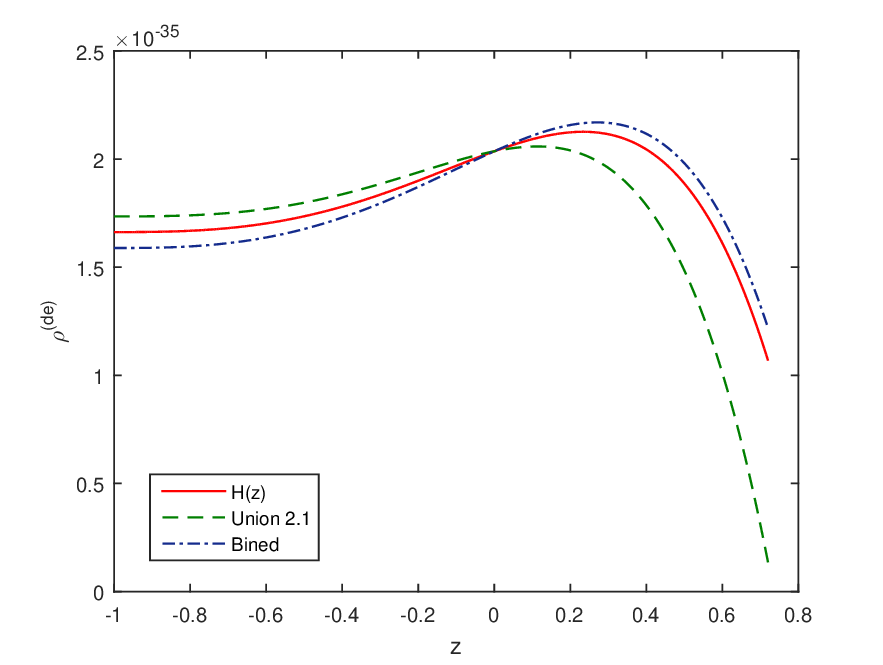}
	b.\includegraphics[width=8cm,height=7cm,angle=0]{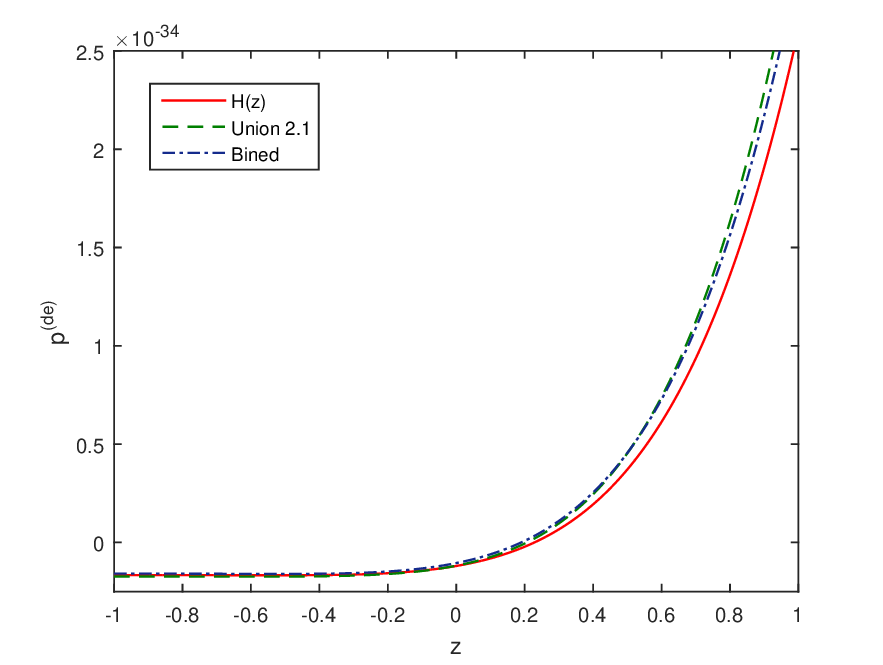}
	\caption{Behaviour of Dark energy $\rho^{(de)}$ and pressure $p^{(de)}$ over $z$, respectively.}
\end{figure}

\subsection*{Dark energy EoS parameter}

The EoS parameter $\omega^{(de)}$ for dark energy factor is defined as
\begin{equation}\label{eq54}
	\omega^{(de)}=\frac{p^{(de)}}{\rho^{(de)}}
\end{equation}
where $\rho^{(de)}$ and $p^{(de)}$ are given by Eqs.~\eqref{eq43} and \eqref{eq44}, respectively. The geometrical evolution of dark energy EoS $\omega^{(de)}$ over redshift $z$ is shown in Figure 5a. From Figure 5a, one can observe that the values of $\omega^{(de)}$ varies as $(-1<\omega^{(de)}<2)$ over the redshift $(-1\le z \le 0.6)$, and it tends to cosmological constant value ($\omega=-1$) as $z\to-1$ during its evolution. We can easily found that the present model shows various stages of dark energy models like matter dominated ($\omega=0$), quintessence $\omega>-1$, cosmological constant $\omega=-1$ dark energy models. We can see that as $z\to-1$, the EoS $\omega^{(de)}\to-1$ along all three datasets, that shows the tendency of the model to $\Lambda$CDM model at late-time universe.\\

\begin{figure}[H]
	\centering
	a.\includegraphics[width=8cm,height=7cm,angle=0]{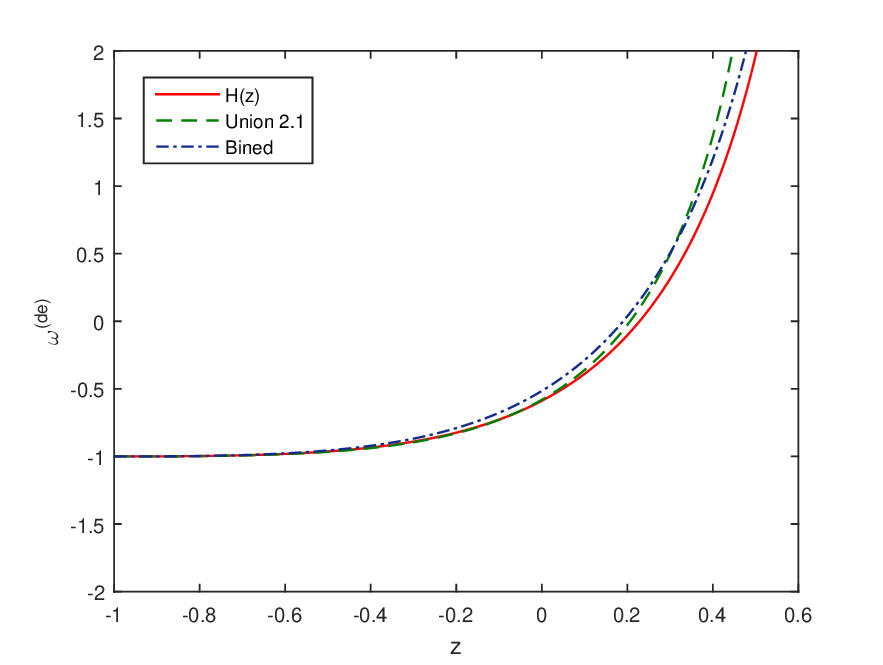}
	b.\includegraphics[width=8cm,height=7cm,angle=0]{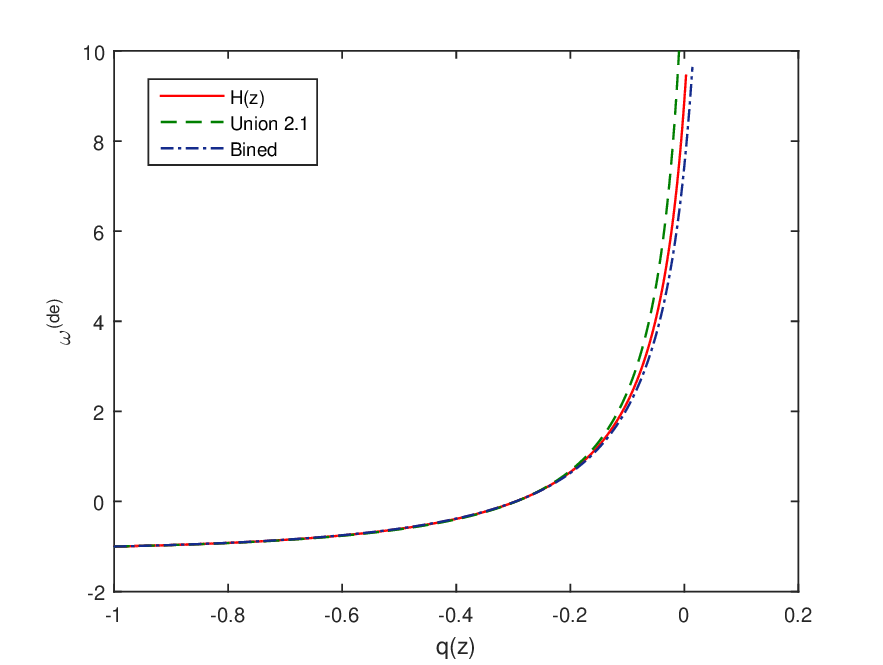}
	\caption{Evolution of dark energy EoS parameter $\omega^{(de)}$ versus $z$ and $\omega^{(de)}$ versus deceleration parameter $q(z)$, respectively.}
\end{figure}
Figure 5b represents the evolution of EoS parameter $\omega^{(de)}$ over deceleration parameter $q(z)$ and we observe that as $q\to-1$, the EoS $\omega^{(de)}\to-1$. The Figure 5b shows the evolution of $\omega^{(de)}$ in various phases of the expanding universe. The estimated present values of dark energy EoS parameter $\omega^{(de)}=-0.5885, -0.5811, -0.5157$, respectively along three datasets $H(z)$, Union $2.1$, and Bined data. Thus, the current values of $\omega^{(de)}$ lies in the range of $(-\frac{1}{3}, -1)$ that suggests our universe model is currently in quintessence phase of evolution. From equation \eqref{eq47}, we can obtain the effective EoS parameter $\omega^{eff}$ as
\begin{equation}\label{eq55}
	\omega^{eff}=-1+\frac{2}{3}\frac{nc_{1} (1+z)^{n}}{2[c_{1}(1+z)^{n}+c_{2}]},
\end{equation}
\begin{figure}[H]
	\centering
	\includegraphics[width=10cm,height=8cm,angle=0]{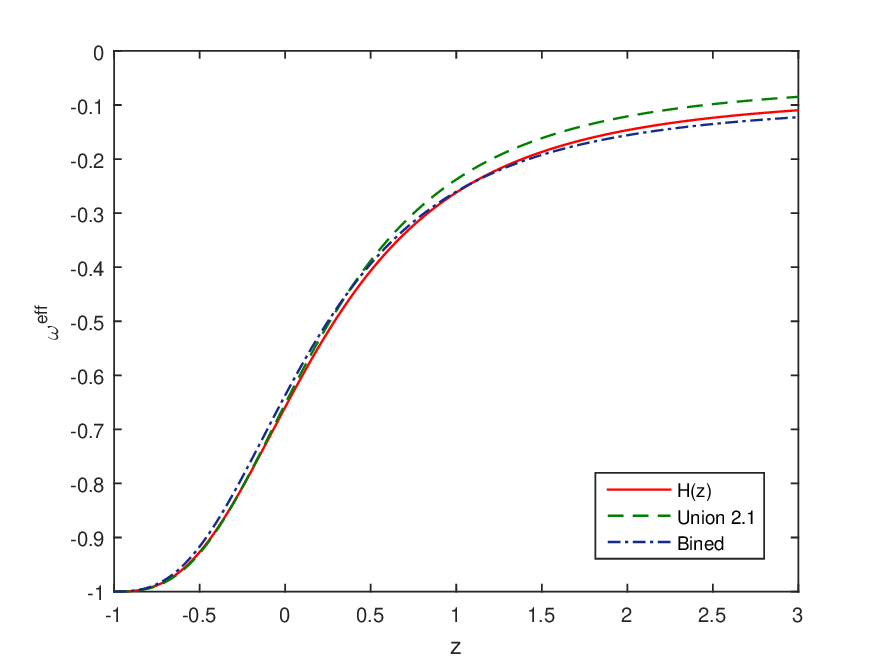}
	\caption{Evolution of effective EoS parameter $\omega^{eff}$ over $z$.}
\end{figure}
The geometrical evolution of the effective EoS parameter $\omega^{eff}$ is shown in figure 6 and we see that as $z\to-1$, then $\omega^{eff}\to-1$ that depicted the tendency of the model towards $\Lambda$CDM model. Also, the estimated present values of effective EoS parameter is found as $\omega^{eff}=-0.6591, -0.6528, -0.6369$ along three observational datasets respectively, that represents the consistency of the model with observational datasets. Also, we can find the value of EoS parameter $\omega^{(m)}$ for matter source as $\omega^{eff}=\omega^{(m)}+\omega^{(de)}$ and this implies $\omega^{(m)}=-0.0706, -0.0717, -0.1212$, respectively for three observational datasets.
\subsection*{Om diagnostic analysis}
The way the Om diagnostic function behaves \cite{ref53} makes it easier to categorize ideas about cosmic dark energy. The Om diagnostic function for a spatially homogeneous world is given as

\begin{equation}\label{eq56}
	Om(z)=\frac{\left(\frac{H(z)}{H_{0}}\right)^{2}-1}{(1+z)^{3}-1},
\end{equation}
where $H_{0}$ represents the current Hubble parameter $H(z)$ value as given in Eq.~\eqref{eq39}. Quintessence motion is indicated by a negative slope of $Om(z)$, whereas phantom motion is indicated by a positive slope. The constant $Om(z)$ stands in for the $\Lambda$CDM model.\\

Using Eq.~\eqref{eq39} in \eqref{eq56}, we obtain
\begin{equation}\label{eq57}
	Om(z)=  \frac{[c_{1}(1+z)^{n}+c_{2}]-1}{(1+z)^{3}-1}
\end{equation}

Figure 7 depicts the geometric development of the $Om(z)$ function, while Eq.~\eqref{eq57} provides its mathematical expression. According to Figure 7, the evolution of the Om diagnostic function has negative slopes for all three datasets, indicating that the model's current state is essential. Its trend to a constant value also indicates the $\Lambda$CDM dark energy model. This demonstrates the accuracy and usefulness of our model.

\begin{figure}[H]
	\centering
	\includegraphics[width=10cm,height=8cm,angle=0]{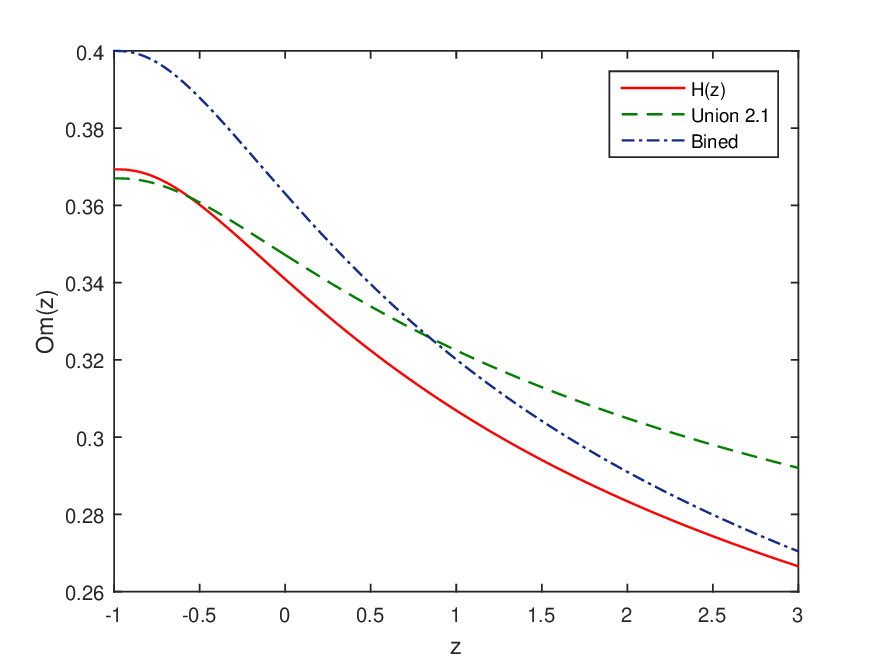}
	\caption{Evolution of Om diagnostic function $Om(z)$ over $z$.}
\end{figure}

\subsection*{Age of the universe}

The age of the cosmic universe is obtained by
\begin{equation}\label{eq58}
	t_{0}-t=\int_{0}^{z}\frac{dz}{(1+z)H(z)}
\end{equation}
where $H(z)$ is given by Eq.~\eqref{eq39}. Using this in \eqref{eq58}, we have
\begin{equation}\label{eq59}
	H_{0}(t_{0}-t)=\lim_{z\to\infty}\int_{0}^{z}\frac{dz}{(1+z)[\lambda(1+z)^{n}+k]^{\frac{1}{2}}}
\end{equation}

We can find that as $z\to\infty$, $H_{0}(t_{0}-t)$ tends to a constant value that represents the cosmic age of the universe, $H_{0}(t_{0}-t)\to H_{0}t_{0}=0.98458, 0.96240, 0.97831$, respectively, along three datasets. The present cosmic age of the universe is estimated as $t_{0}=14.16, 13.84, 14.07$ Gyrs, respectively along three observational datasets, which are very closed to observational estimated values.
\section{Conclusions}
		In the recently suggested modified non-metricity gravity theory with boundary term in a flat FLRW spacetime universe, dark energy scenarios of cosmological models are examined in this study. An arbitrary function, $f(Q,C)=Q+\alpha C^{2}$, has been taken into consideration, where $Q$ is the non-metricity scalar, $C$ is the boundary term denoted by $C=\mathring{R}-Q$, and $\alpha$ is the model parameter, for the action that is quadratic in $C$. The Hubble function $H(z)=H_{0}[c_{1}(1+z)^{n}+c_{2}]^{\frac{1}{2}}$, where $H_{0}$ is the current value of the Hubble constant and $n, c_{1}$ and $c_{2}$ are arbitrary parameters with $c_{1}+c_{2}=1$, has been used to examine the dark energy characteristics of the model. We discovered a transit phase expanding universe model that is both decelerated in the past and accelerated in the present, and we discovered that the dark energy equation of state (EoS) $\omega^{(de)}$ behaves as $(-1\le\omega^{(de)}<2)$. The Om diagnostic analysis reveals the quintessence behavior in the present and the cosmological constant scenario in the late-time universe. Finally, we calculated the universe's current age, which was found to be quite similar to recent data. The estimated values of various parameters are shown in below Table 4.
\begin{table}[H]
	\centering
	\begin{tabular}{|c|c|c|c|}
		\hline
		
		Parameter      & $H(z)$ data         & Union 2.1 data        &  Bined data\\
		\hline
		$c_{1}$      & $0.36934\pm0.04197$ & $0.3670\pm0.03877$    & $0.40\pm0.032996$\\
		$n$            & $2.76902\pm0.08594$ & $2.83802\pm0.65173$   & $2.72294\pm1.34523$\\
		$c_{2}$            & $0.63066$           & $0.6330$              & $0.60$\\
		$H_{0}$        & $68\pm0.84653$      & $--$                  & $--$\\
		$q_{0}$        & $-0.4886$           & $-0.4792$             & $-0.4534$\\
		$t_{0}$        & $14.16$ Gyrs        & $13.84$ Gyrs          & $14.07$ Gyrs\\
		$z_{t}$        & $0.7132$            & $0.6463$              & $0.6864$\\
		$\omega^{(de)}$& $-0.5885$           & $-0.5811$             & $-0.5157$\\
		$\omega^{eff}$ & $-0.6591$           & $-0.6528$             & $-0.6369$\\
		$\alpha$       & $2.7291\times10^{35}$          & $2.8272\times10^{35}$            & $2.8814\times10^{35}$\\
		\hline
	\end{tabular}
	\caption{Estimated values of various parameters.}\label{T4}
\end{table}
The main features of the model are as follows:
\begin{itemize}
	\item We have obtained a transit phase accelerating universe model with present value of deceleration parameter $q_{0}=-0.4886, -0.4792, -0.4534$ along three observational datasets, respectively and found that $q\to-1$ at late-time universe.
	\item We have estimated the transition redshift $z_{t}=0.7132, 0.6463, 0.6864$ along three datasets, respectively which very closed to recent observed values \cite{ref50}-\cite{ref52}.
	\item We have found that the boundary term behaves just like dark energy candidate, cause of acceleration in expanding universe.
	\item We have investigated the dynamical behaviour of dark energy parameter $\omega^{(de)}$ and found as quintessence region at present $(\omega^{(de)}>-1)$ and tends to cosmological constant value $(\omega^{(de)}=-1)$ at late-time universe.
	\item The Om diagnostic analysis shows the behaviour of quintessence dark energy at present and tends to $\Lambda$CDM dark energy model at late-time universe.
	\item The estimated present age of the universe $t_{0}=14.16, 13.84, 14.07$ Gyrs along three datasets, respectively and the present value of Hubble constant $H_{0}=68\pm0.84653~Km/s/Mpc$ are in good agreement with recent observed values.
\end{itemize}
Thus, our derived model is interesting to readers and researchers for more investigations.
\section*{Acknowledgments}
 The author (A. Pradhan ) is grateful for the assistance and facilities provided by the University of Zululand, South Africa during a visit where a part of this article was completed.

\section{Declarations}
\subsection*{Funding and/or Conflicts of interests/Competing interests}
The authors of this article have no conflict of interests. The authors have no competing interests to declare that are relevant to the content of this article. The author did not receive support from any organization for the submitted work.


\begin{thebibliography}{}
	\bibitem {ref1}
	S. Perlmutter \textit{et al.}, Discovery of a supernova explosion at half the age of the Universe, \textit{ Nature} \textbf{391} 51 (1998); A. G. Riess \textit{et al.}, Observational evidence from supernovae for an accelerating universe and a cosmological constant, \textit{Astron. J.} \textbf{116} 1009 (1998).
	\bibitem{ref2}
	D. J. Eisenstein, W. Hu, and M. Tegmark, Cosmic Complementarity: $H_{0}$ and $\Omega_{m}$ from Combining Cosmic Microwave Background Experiments and Redshift Surveys, \textit{Astrophys. J.} \textbf{504} L57 (1998).
	\bibitem{ref3}
	N. Aghanim \textit{et al.} (Planck Collaboration), Planck 2018 results. VI. Cosmological parameters, \textit{Astron. Astrophys.} \textbf{641} A6 (2020).
	\bibitem{ref4}
	B. S. Haridasu, \textit{et al.}, Strong evidence for an accelerating universe, \textit{Astron. Astrophys.} \textbf{600} L1 (2017).
	\bibitem{ref5}
	V. Sahni and A. Starobinsky, The case for a positive cosmological $\Lambda$-term, \textit{Int. J. Mod. Phys. D} \textbf{9} 373 (2000); S. M. Carroll, The cosmological constant, \textit{Living Rev. Rel.} \textbf{4} 1 (2001); P. J. E. Peebles and B. Ratra, The cosmological constant and dark energy, \textit{Rev. Mod. Phys.} \textbf{75} 559 (2003); E. J. Copeland, M. Sami and S. Tsujikawa, Dynamics of dark energy, \textit{Int. J. Mod. Phys. D} \textbf{15} 1753 (2006).
	\bibitem{ref6}
	S. Weinberg, The cosmological constant problem, \textit{Rev. Mod. Phys.} \textbf{61} 1 (1989); T. Padmanabhan, Cosmological constant-the weight of the vacuum, \textit{Phys. Rept.} \textbf{380} 235 (2003).
	
	\bibitem{ref7}
	S. Capozziello, R. D'Agostino and O. Luongo, Extended gravity cosmography, \textit{Int. J. Mod. Phys. D} \textbf{28} 1930016 (2019); S. Nojiri and S. D. Odintsov, Unified cosmic history in modified gravity: from $F(R)$ theory to Lorentz non-invariant models, \textit{Phys. Rept.} \textbf{505} 59 (2011); S. Capozziello, M. De Laurentis, Extended theories of gravity, \textit{Phys. Rept.} \textbf{509} 167 (2011); K. Bamba, S. Capozziello, S. Nojiri and S. D. Odintsov, Dark energy cosmology: the equivalent description via different theoretical models and cosmography tests, \textit{Astrophys. Space Sci.} \textbf{342} 155 (2012).
	\bibitem{ref8}
	A. De Felice, S. Tsujikawa, $f(R)$ Theories, \textit{Living Rev. Rel.} \textbf{13} 3 (2010); S. Nojiri, S. D. Odintsov, V. K. Oikonomou, Modified gravity theories on a nutshell: Inflation, bounce and late-time evolution, \textit{Phys. Rept.} \textbf{692} 1 (2017).
	\bibitem{ref9}
	G. R. Bengochea and R. Ferraro, Dark torsion as the cosmic speed-up, \textit{Phys. Rev. D} \textbf{79} 124019 (2009); E. Linder, Einstein's other gravity and the acceleration of the universe, \textit{Phys. Rev. D} \textbf{82} 109902 (2010); Y. F. Cai, S. Capozziello, M. De Laurentis, E. N. Saridakis, $f(T)$ teleparallel gravity and cosmology, \textit{Rept. Prog. Phys.} \textbf{79} 106901 (2016); R. D'Agostino, O. Luongo, Growth of matter perturbations in nonminimal teleparallel dark energy, \textit{Phys. Rev. D} \textbf{98} 124013 (2018).
	\bibitem{ref10}
	R. D'Agostino, R. C. Nunes, Measurements of $H_{0}$ in modified gravity theories: The role of lensed quasars in the late-time Universe, \textit{Phys. Rev. D} \textbf{101} 103505 (2020); A. Bonilla, R. D'Agostino, R. C. Nunes, J. C. N. de Araujo, Forecasts on the speed of gravitational waves at high z, \textit{J. Cosm. Astrop. Phys.} \textbf{03} 015 (2020); R. D'Agostino, R. C. Nunes, Probing observational bounds on scalar-tensor theories from standard sirens, \textit{Phys. Rev. D} \textbf{100} 044041 (2019).
	\bibitem{ref11}
	J. Beltr\'{a}n Jim\'{e}nez, L. Heisenberg, T. Koivisto, Coincident general relativity, \textit{Phys. Rev. D} \textbf{98}, 044048 (2018); J. Beltran Jimenez, L. Heisenberg, T. S. Koivisto and S. Pekar, Cosmology in $f(Q)$ geometry, \textit{Phys. Rev. D} \textbf{101} (2020) 103507.; E. Saridakis \textit{et al.} Modified Gravity and Cosmology, Springer, Cham, Switzerland (2021).
	\bibitem{ref12}
	P. J. Peebles and B. Ratra, Cosmology with a Time-Variable ``Cosmological Constant'', \textit{Astrophys. J.} \textbf{325} 1220 (1988); R. R. Caldwell, R. Dave and P. J. Steinhardt, Cosmological imprint of an energy component with general equation of state, \textit{Phys. Rev. Lett.} \textbf{80} 1582 (1998); R. D'Agostino, Holographic dark energy from nonadditive entropy: cosmological perturbations and observational constraints, \textit{Phys. Rev. D} \textbf{99} 103524 (2019); R. D'Agostino and O. Luongo, Cosmological viability of a double field unified model from warm inflation, (2021) e-Print:2112.12816 [astro-ph.CO].
	\bibitem{ref13}
	S. Capozziello, R. D'Agostino and O. Luongo, Cosmic acceleration from a single fluid description, \textit{Phys. Dark Univ.} \textbf{20} 1 (2018); K. Boshkayev, R. D'Agostino and O. Luongo, Extended logotropic fluids as unified dark energy models, \textit{Eur. Phys. J. C} \textbf{79} 332 (2019); S. Capozziello, R. D'Agostino, R. Giamb\`{o}, O. Luongo, Effective field description of the Anton-Schmidt cosmic fluid, \textit{Phys. Rev. D} \textbf{99} 023532 (2019).
	\bibitem{ref14}
	F. Bajardi, D. Vernieri, S. Capozziello, Bouncing cosmology in $f(Q)$ symmetric teleparallel gravity, \textit{Eur. Phys. J. Plus} \textbf{135} 912 (2020); B. J. Barros, T. Barreiro, T. Koivisto, N. J. Nunes, Testing $F(Q)$ gravity with redshift space distortions, \textit{Phys. Dark Univ.} \textbf{30} 100616 (2020); A. Narawade, L. Pati, B. Mishra, S. K. Tripathy, Dynamical system analysis for accelerating models in non-metricity f (Q) gravity, arXiv:2203.14121.
	\bibitem{ref15}
	I. Ayuso, R. Lazkoz, V. Salzano, Observational constraints on cosmological solutions of $f(Q)$ theories, \textit{Phys. Rev. D} \textbf{103} 063505 (2021); N. Frusciante, Signatures of $f(Q)$-gravity in cosmology, \textit{Phys. Rev. D} \textbf{103}, 044021 (2021); F. K. Anagnostopoulos, S. Basilakos, E. N. Saridakis, First evidence that non-metricity $f(Q)$ gravity could challenge $\Lambda$CDM, \textit{Phys. Lett. B} \textbf{822} 136634 (2021).
	\bibitem{ref16}
	S. Mandal, D. Wang, P. K. Sahoo, Cosmography in $f(Q)$ gravity, \textit{Phys. Rev. D} \textbf{102}, 124029 (2020); G. Mustafa, Z. Hassan, P. H. R. S. Moraes, P. K. Sahoo, Wormhole solutions in symmetric teleparallel gravity, \textit{Phys. Lett. B} \textbf{821}, 136612 (2021); R. Solanki, A. De, P. K. Sahoo, Complete dark energy scenario in $f(Q)$ gravity, \textit{Phys. Dark Univ.} \textbf{36} 100996 (2022);	A. Pradhan, D. C. Maurya and A. Dixit, Dark energy nature of viscus universe in $f(Q)$-gravity with observational constraints, \textit{Int. J. Geom. Meth. Mod. Phys.} \textbf{18}, 2150124 (2021);	A. Dixit, D. C. Maurya and A. Pradhan, Phantom dark energy nature of bulk-viscosity universe in modified $f(Q)$-gravity, \textit{Inter. J. Geom. Meth. Mod. Phys.} \textbf{19} (2022) 2250198-581; A. Pradhan, A. Dixit and D. C. Maurya, Quintessence Behavior of an Anisotropic Bulk Viscous Cosmological Model in Modified $f(Q)$-Gravity, \textit{Symmetry} \textbf{14} (2022) 2630; 	M. Koussour, S. H. Shekh, M. Bennai, Anisotropic $f(Q)$ gravity model with bulk viscosity, \textit{https://doi.org/10.48550/arXiv.2203.10954}.
	\bibitem {ref17}
	T. Harko \textit{et al.}, Coupling matter in modified $f(Q)$ gravity, \textit{Phys. Rev. D} \textbf{98} (2018) 084043. arXiv:1806.10437 [gr-qc].
	\bibitem {ref18}
	S. Capozziello and R. D'Agostino, Model-independent reconstruction of $f(Q)$ non-metric gravity, \textit{Phys. Lett. B} \textbf{832} (2022) 137229.
	\bibitem {ref19}
	L. J\"{a}rv, M. R\"{u}nkla, M. Saal, O. Vilson, Nonmetricity formulation of general relativity and its scalar-tensor extension, \textit{Phys. Rev. D} \textbf{97}, 124025 (2018).
	\bibitem {ref20}
	F. W. Hehl, G. D. Kerlick, P. van der Heyde, General relativity with spin and torsion: Foundations and prospects, \textit{Zeitschrift f\"{u}r Naturforschung A} \textbf{31} 111 (1976)
	\bibitem {ref21}
	D. Zhao, Covariant formulation of $f(Q)$ theory, \textit{Eur. Phys. J. C} \textbf{82} 303 (2022).
	\bibitem {ref22}
	Y. Xu, G. Li, T. Harko and S. D. Liang, $f(Q,T)$ gravity, \textit{Eur. Phys. J. C} \textbf{79} 708 (2019).
	\bibitem{ref23}
	Avik De, Tee-How Loo, and E. N. Saridakis, Non-metricity with boundary terms: $f(Q,C)$ gravity and cosmology, (2023) arxiv:2308.00652v1v1 [gr-qc].
	\bibitem{ref24}
	S. Capozziello, V. D. Falco, and C. Ferrara, The role of the boundary term in $f(Q,B)$ symmetric teleparallel gravity, (2023) arxiv:2307.13280v1 [gr-qc].
	\bibitem {ref25}
    D.C. Maurya, Transit cosmological model with specific Hubble parameter in $F(R, T)$ gravity, \textit{New Astronomy} \textbf{77} (2020) 101355.
	\bibitem {ref26}
    E. J. Copeland, \textit{et al.}, Dynamics of dark energy, \textit{Int. J. Mod. Phys. D} \textbf{15} (2006) 1753.
	\bibitem {ref27}
	S. Agarwal, R. K. Pandey, A. Pradhan, LRS Bianchi type II perfect fluid cosmological models in normal gauge for Lyra's manifold, \textit{Int. J. Theor. Phys.} {\bf 50} (2011) 296-307.
	\bibitem {ref28}
	A. Pradhan, S. Agarwal, G. P. Singh, LRS Bianchi type-I universe in Barber's second self creation theory, \textit{Int. J. Theor. Phys.} {\bf 48} (2009) 158-166.
	\bibitem {ref29}
	E. Macaulay, \textit{et al.}, First cosmological results using Type Ia supernovae from the dark energy survey: measurement of the Hubble constant,	\textit{Mon. Not. R. Astro. Soc.} {\bf 486} (2019) 2184-2196.
	\bibitem {ref30}
	C. Zhang, \textit{et al.}, Four new observational $H(z)$ data from luminous red galaxies in the sloan digital sky survey data release seven,	\textit{Res. Astron. Astrophys.} {\bf 14} (2014) 1221-1233.
	\bibitem {ref31}
	D. Stern, \textit{et al.}, Cosmic chronometers: constraining the equation of state of dark energy.I: $H(z)$ measurements,
	\textit{J. Cosmol. Astropart. Phys.} {\bf 1002} (2010) 008.
	\bibitem {ref32}
	E. G. Naga, \textit{et al.}, Clustering of luminous red galaxies-IV: Baryon acoustic peak in the line-of-sight direction and a direct measurement of $H(z)$, \textit{Mon. Not. R. Astro. Soc.} {\bf 399} (2009) 1663-1680.
	\bibitem {ref33}
	D. H Chauang, Y. Wang, Modelling the anisotropic two-point galaxy correlation function on small scales and single-probe measurements of $H(z)$, $D_{A}(z)$ and $f(z)$, $\sigma_{8}(z)$ from the sloan digital sky survey DR7 luminous red galaxies, \textit{Mon. Not. R. Astro. Soc.} {\bf 435} (2013) 255-262.
	\bibitem {ref34}
	S. Alam, \textit{et al.}, The clustering of galaxies in the completed SDSS-III Baryon Oscillation Spectroscopic Survey: cosmological analysis of the DR12 galaxy sample, \textit{Mon. Not. R. Astron. Soc.} {\bf 470} (2017) 2617.
	\bibitem {ref35}
	A. L. Ratsimbazafy, \textit{et al.}, Age-dating luminous red galaxies observed with the Southern African Large Telescope,
	\textit{Mon. Not. R. Astron. Soc.} {\bf 467} (2017) 3239.
	\bibitem {ref36}
	L. Anderson, \textit{et al.}, The clustering of galaxies in the SDSS-III Baryon oscillation Spectro-scopic Survey: baryon acoustic oscillations in the data releases 10 and 11 galaxy samples, \textit{Mon. Not. R. Astron. Soc.} {\bf 441} (2014) 24.
	\bibitem {ref37}
	M. Moresco, Raising the bar: new constraints on the Hubble parameter with cosmic chronometers at z $\equiv$ 2, \textit{Mon. Not. R. Astron. Soc.} {\bf 450} (2015) L16.
	\bibitem {ref38}
	N. G. Busa, \textit{et al.}, Baryon acoustic oscillations in the Ly$\alpha$ forest of BOSS quasars, \textit{Astron. $\&$ Astrophys} {\bf 552} (2013) A96.
	\bibitem {ref39}
	M. Moresco, \textit{et al.}, Improved constraints on the expansion rate of the Universe up to $z\sim 1.1$ from the spectroscopic evolution of cosmic chronometers, \textit{J. Cosmol. Astropart. Phys.} {\bf 2012} (2012) 006.
	\bibitem {ref40}
	J. Simon, L. Verde, R. Jimenez, Constraints on the redshift dependence of the dark energy potential, \textit{Phys. Rev. D} {\bf 71}	(2005) 123001.
	\bibitem {ref41}
	M. Moresco \textit{et al.}, A 6 $\%$ measurement of the Hubble parameter at z $\sim 0.45$ direct evidence of the epoch of cosmic re-acceleration,	\textit{J. Cosmol. Astropart. Phys.}  {\bf 05} (2016) 014.
	\bibitem {ref42}
	G. F. R. Ellis, M. A. H. MacCallum, A class of homogeneous cosmological models, \textit{Commun. Math. Phys.} {\bf 12} (1969) 108.
	\bibitem{ref43}
	Planck Collaboration, Aghanim N, Akrami Y et al (2020) Planck 2018 results. VI. Cosmological
	parameters. A \& A 641:A6. https://doi.org/10.1051/0004-6361/201833910. arXiv:1807.06209 [astroph.CO]
	\bibitem{ref44}
	Riess AG, Casertano S, Yuan W et al (2021) Cosmic distances calibrated to $1\%$ precision with Gaia
	EDR3 parallaxes and Hubble Space Telescope photometry of 75 Milky Way Cepheids Confirm Tension with KCDM. ApJ 908(1):L6. https://doi.org/10.3847/20418213/abdbaf. arXiv:2012.08534
	[astro-ph.CO]
	\bibitem{ref45}
	G. K. Goswami, R. N. Dewangan and A. K. Yadav, Anisotropic universe with magnetized dark energy, {\it Astrophys. Space Sci.} {\bf 361}, 119 (2016).
	\bibitem{ref46}
	G. K. Goswami, R. N. Dewangan, A. K. Yadav and A. Pradhan, Anisotropic string cosmological models in Heckmann-Schucking space-time, {\it Astrophys. Space Sci.} {\bf 361}, 47 (2016).
	\bibitem {ref47}
	A. K. Camlibel, I. Semiz and M. A. Feyizoglu, Pantheon update on a model-independent analysis of cosmological supernova data, \textit{Class. Quantum Grav.} \textbf{37} (2020) 235001.
	\bibitem {ref48}
	D. M. Scolnic \textit{et al.}, The complete light-curve sample of spectroscopically confirmed SNe Ia from Pan$-$STARRS1 and cosmological constraints from the combined pantheon sample, \textit{Astrophys. J.} \textbf{859} (2018) 101.
	\bibitem{ref49}
	Suzuki, N.; Rubin, D.; Lidman, C.; Aldering, G.; Amanullah, R.; Barbary, K.; Barrientos, L. F.; Botyanszki, J.; Brodwin, M.; Connolly, N.; Dawson, K. S. The Hubble space telescope cluster supernova survey. V. Improving the dark-energy constraints above $z > 1$ and building an early-type-hosted supernova sample, \textit{Astrophys. J.} {\bf 2012}, \textit{746}, 85.
	\bibitem{ref50}
	Spurnova Serach Team collaboration (A. G. Riess {\it et al.}), New hubble space telescope discoveries of type Ia supernovae at $z > 1$:
	narrowing constraints on the early behavior of dark energy, {\it Astrophys. J.} {\bf 659} (2007) 98.
	\bibitem {ref51}
	P. Astier \textit{et al.}, The Supernova Legacy Survey: measurement of, and w from the first year data set, \textit{Astron. Astrophys.} {\bf 447} (2006) 31-48.
	\bibitem {ref52}
	J.A.S. Lima, J.F. Jesus, R.C. Santos, M.S.S. Gill, Is the transition redshift a new cosmological number?, arXiv:1205.4688v2[astro-ph.CO], (2012).
	\bibitem {ref53}
	V. Sahni, A. Shafieloo, A. A. Starobinsky, Two new diagnostics of dark energy, \textit{Phys. Rev. D} \textbf{78} (2008) 103502.
\end{thebibliography}
\end{document}